\documentclass[twocolumn]{aastex63}

\newcommand{\src}{FRB\,180309}
\newcommand{\xsrc}{J2124-3358}

\newcommand{\dmunits}{{\rm pc\,cm\,$^{-3}$}}
\usepackage{color}
\usepackage{graphicx, epstopdf} 
\usepackage{gensymb}

\newcommand{\dmcosmic}{DM$_{\rm cosmic}$}

\newcommand{\hi}{H~{\sc i}}

\graphicspath{{./}{figures/}}

\received{-}
\revised{-}
\accepted{-}
\submitjournal{ApJ}

\shorttitle{\src\ Follow-up}
\shortauthors{Aggarwal et al.}

\begin{document}

\title{Multi-wavelength follow-up of \src}

\correspondingauthor{Kshitij Aggarwal}
\email{ka0064@mix.wvu.edu}

\author[0000-0002-2059-0525]{Kshitij Aggarwal}
\affil{West Virginia University, Department of Physics and Astronomy, P. O. Box 6315, Morgantown, WV, USA}
\affil{Center for Gravitational Waves and Cosmology, West Virginia University, Chestnut Ridge Research Building, Morgantown, WV, USA}

\author[0000-0003-4052-7838]{Sarah Burke-Spolaor}
\affil{West Virginia University, Department of Physics and Astronomy, P. O. Box 6315, Morgantown, WV, USA}
\affil{Center for Gravitational Waves and Cosmology, West Virginia University, Chestnut Ridge Research Building, Morgantown, WV, USA}

\author[0000-0002-1883-4252]{Nicolas Tejos}
\affiliation{Instituto de F\'isica, Pontificia Universidad Cat\'olica de Valpara\'iso, Casilla 4059, Valpara\'iso, Chile}

\author[0000-0003-0006-0188]{Giuliano Pignata}
\affiliation{Departamento de Ciencias Fisicas, Universidad Andres Bello, Avda.  Republica 252, Santiago, Chile}
\affiliation{Millennium Institute of Astrophysics (MAS), Nuncio Monsenor Sotero Sanz 100, Providencia,Santiago, Chile}

\author[0000-0002-7738-6875]{J. Xavier Prochaska}
\affil{University of California Observatories-Lick Observatory, University of California, 1156 High Street, Santa Cruz, CA95064, USA}
\affil{Kavli Institute for the Physics and Mathematics of the Universe (WIP), 5-1-5 Kashiwanoha, Kashiwa, 277-8583, Japan}

\author[0000-0002-7252-5485]{Vikram Ravi}
\affil{Cahill Center for Astrophysics, California Institute of Technology, Pasadena, CA 91125, USA}

\author[0000-0003-4810-7803]{Jane F. Kaczmarek}
\affil{CSIRO Astronomy and Space Science, Australia Telescope National Facility, Box 76, Epping, NSW 1710, Australia}

\author{Stefan Os{\l}owski}
\affil{Gravitational Wave Data Centre, Swinburne University of Technology, P.O. Box 218, Hawthorn, VIC 3122, Australia}
\affil{Centre for Astrophysics and Supercomputing, Swinburne University of Technology, P.O. Box 218, Hawthorn, VIC 3122, Australia}

\begin{abstract}
We report on the results of multi-wavelength follow-up observations with Gemini, VLA, and ATCA, to search for a host galaxy and any persistent radio emission associated with \src. This FRB is among the most luminous FRB detections to date, with a luminosity of $>8.7\times 10^{32}$\,erg\,Hz$^{-1}$
at the dispersion-based redshift upper limit of 0.32. We used the high-significance detection of \src\ with the Parkes Telescope and a beam model of the Parkes Multibeam Receiver to improve the localization of the FRB to a region spanning approximately $\sim2'\times2'$. We aimed to seek bright galaxies within this region to determine the strongest candidates as the originator of this highly luminous FRB. We identified optical sources within the localization region above our $r$-band magnitude limit of 24.27, fourteen of which have photometric redshifts whose fitted mean is consistent with the redshift upper limit ($z<0.32$) of our FRB.
Two of these galaxies are coincident with marginally detected ``persistent'' radio sources of flux density 24.3\,$\mu$Jy\,beam$^{-1}$ and 22.1\,$\mu$Jy\,beam$^{-1}$ respectively. Our redshift-dependent limit on the luminosity of any associated persistent radio source is comparable to the luminosity limits for other localized FRBs. We analyze several properties of the candidate hosts we identified, including chance association probability, redshift, and presence of radio emission, however it remains possible that any of these galaxies could be the host of this FRB. Follow-up spectroscopy on these objects to explore their H$\alpha$ emission and ionization contents, as well as to obtain more precisely measured redshifts, may be able to isolate a single host for this luminous FRB.
%Far from the FRB localization region, we discovered a complex X-shaped source in our VLA image, which appears similar to the S-shaped radio galaxies reported in the literature.  

\end{abstract}


% radio continuum: general --- methods: observational --- catalogs --- instrumentation: interferometers}

\section{Introduction} \label{sec:intro}
Fast Radio Bursts (FRBs) are bright, millisecond-duration radio transients of unknown origin and are characterized by dispersion measures (DM) that are much higher than the Milky Way's contribution in a given direction \citep{lorimer07}. 
%They were first seen in data from Parkes Telescope \citep{lorimer07}, but have subsequently been detected at other radio telescopes as well \citep{masui15,thronton2013,amiri2019}. More than 60 
Hundreds of such bursts have been seen so far, most of which have been one-off events, while some %including two FRB\,121102 \citep{spitler16} and FRB180814.J0422+73\citep{r2amiri2019} 
have been seen to repeat \citep[][]{Petroff2016}.

In only the past three years, a number of FRBs have been localized using 
%JXP -- swapped the order, as is sensible for many reasons
the Australian Square Kilometre Array Pathfinder (ASKAP), 
Deep Synoptic Array (DSA), 
the European VLBI Network (EVN), and the {\sc realfast}/Very Large Array (VLA) experiment \citep{bannister2019, Ravi+2019, Prochaska2019Sci, marcote20, law2020, Macquart2020}. 
%Due to its repeating nature, follow-up observations of
The repetitions of the first known repeating FRB, FRB\,121102, led to its sub-arcsecond localization \citep{chatterjee17}. It was further associated with a low-metallicity faint dwarf galaxy at $z=0.19$ \citep{tendulkar2017}.  It is also coincident with a highly compact (projected size $<$\,0.7~pc), persistent (non-bursty) radio source of flux density S$_{\text{2 GHz}} \sim 150 \mu$Jy, which has a distinct spectrum that is relatively flat out to around 8~GHz, and declines above that cutoff \citep{chatterjee17, marcote17}. While the same report indicated that chance coincidence of this emission's colocation with the FRB is exceedingly small ($P<10^{-5}$), it remains possible that the emission colocated with FRB\,121102 is simply a red herring, and not specifically caused by or related to the FRB.

The properties of the subsequently localized bursts and their hosts are markedly different from that of FRB\,121102, ranging from massive elliptical galaxies to luminous spiral galaxies \citep{bannister2019,Ravi+2019, Prochaska2019Sci, bhandari2020, marcote20}. Only one other repeating FRB (FRB\,180916.J0158+65) has been localized \citep{marcote20} to a star-forming region in a massive spiral galaxy. Similar to FRB\,121102, FRB\,180916.J0158+65 is offset from this knot of star forming region \citep[][]{bassa2017, tendulkar2021}. No persistent radio emission has been detected in any of these other localized FRBs. It is to be noted that upper limits on persistent radio emission for some of these FRBs are a few orders of magnitude less than the persistent emission detected for FRB\,121102 \citep[][]{marcote20, Macquart2020}. Thus, it appears that FRBs may originate from a variety of host galaxies and environments \citep[][]{Heintz2020}. Recently, luminous radio bursts have been detected from a Galactic magnetar SGR 1935+2154, providing evidence that some FRBs may be emitted by magnetars at cosmological distances
 and originate from Milky Way like galaxies \citep{Bochenek2020, 1935chime2020, kirsten2020}.   

%\citet{Ravi+2019, bannister2019} reported the detection and localization of FRB 190523 and FRB180924 using Deep Synoptic Array (DSA) and Australian Square Kilometre Array Pathfinder (ASKAP) iterferometers, respectively. The properties of both these bursts and their hosts are markedly different from that of FRB\,121102. 
%Unlike FRB\,121102, neither of these FRBs have been seen to repeat and have been associated with massive luminous galaxies. The host galaxy of FRB190523 also indicates that its progenitor may have been from an old stellar population and is similar to Milky Way. This implies that galaxies like Milway Way, and not just dwarf galaxies, could also host FRBs. Moreover, no persistent radio source was seen for FRB180924, down to a limit of 20$\mu$Jy at 6.5 GHz. 

%Extreme conditions are needed to explain the high luminosities, short durations, and high polarization fractions of FRBs \citep{katz2018}. FRBs show a broad variety of properties in their polarization (ranging from 0--100\% polarized), pulse substructures (unresolved, Gaussian, scattered, and multi-peaked), and repetition rate (at least two are now known to repeat on timescales of days \citep{spitler16,r2amiri2019}, while others have been observed for a few tens to hundreds of hours over several years without further detection \citep{petroff2015, lorimer07}).

Sub-arcsecond localization is generally essential for conclusive identification of an FRB with a counterpart or host galaxy \citep{eftekhari2017,bloom2002}. This is crucial to understanding the progenitors of FRBs and their host environments. But subarcsecond localization is only possible with interferometers with realtively large baselines like VLA, ASKAP and EVN. With single-dish telescopes, the localization is limited by the beam size, which is usually several arcminutes. Multibeam receivers, like the ones at Parkes \citep[][]{Smith1996}, and Focal L-band Array for the Green Bank Telescope \citep[][]{Rajwade2018}, can improve localization if a signal is bright enough to appear in multiple beams, allowing one to fit the relative signal strength in each beam to the side-lobe pattern of the multiple beams. This technique has been used to some success for past bursts; for instance, \citet{ravi2019} localized FRB\,010724 (detected in 3 beams with a 1-bit system) to around 50~arcmin$^2$ with no frequency information or calibration. With an improved 8-bit observing system, \citet{ravi2016} was able to localize FRB\,150807 with two beams to around 9~arcmin$^2$.

\citet{oslowski2019} reported the discovery of four FRBs found with the commensal FRB search system at Parkes Telescope. These FRBs were detected during the Parkes Pulsar Timing Array observations of millisecond pulsars. Of these four, \src\ was the strongest FRB yet detected, with
%with Parkes with a detection S/N of 411. It was 
a fluence ($F>83.5$~Jy\,ms) that was so bright that it saturated the central beam of the Parkes Telescope Multibeam Receiver \citep[][]{Smith1996}. It was detected at a DM of 263.42~\dmunits\ and was narrowest of those 4 FRBs with a width of 0.475~ms. \citet{oslowski2019} estimated its redshift to be $z<0.19$ and calculated the linear and circular polarization fractions to be $L_{\text{f}} = 0.4556 \pm 0.0006$ and $V_{\text{f}} = 0.2433 \pm 0.0005$. They also placed an upper limit of 150~rad\,m\,$^{-2}$ on the modulus of the rotation measure of this FRB. More importantly, it was also detected 
%at much less S/N, 
in six other beams, with a flux fall-off that matches that expected from side-lobe detection from those beams (i.e. this signal was localized on the sky roughly boresight to the pointing position, and is not like the terrestrial peryton signals reported by \citealt{spolaor2011}). Given the seven-beam detection of \src\ and fully calibrated spectra, it is possible to greatly improve the standard 14$' \times 14'$ localization typically provided by Parkes.
%Our improved localization enabled us to do multi-frequency follow-up observations of this FRB, to search for any optical host galaxy counterpart and any persistent radio emission. 

%This FRB has a DM of 263.42 $\pm$ 0.01\dmunits  of which the galactic contribution is 30\dmunits. \citet{oslowski2019} placed a redshift upper limit of 0.187 on it. 

In this paper, we summarize the process for improving the localization of this FRB to approximately $2' \times 2'$. We performed multi-frequency follow-up observations of this error region to search for optical host galaxy candidates and any candidates for persistent radio sources related to the FRB within its DM-based redshift upper limit.
%Follow-up observations of \src\ with Gemini, Very Large Array (VLA) and Australia Telescope Compact Array (ATCA) to identify or put stringent limits on a persistent astrophysical source. 
In \S2 we describe radio and optical follow-up observations. In \S3 we re-examine the likely redshift range of the FRB based on DM-$z$ relationships that have been updated since the publication of \citet{oslowski2019}. We describe the procedure of fitting the beam pattern to obtain a more precise localization in \S4. \S5 presents the results of our observations, and we discuss the implications of our analysis and results in \S6. 

% Estimated fluence of \src\  is $F=76.8$Jy ms.

\section{Observations}\label{sec:data}
    % \subitem Astrometric comparison (including position table). - S
\subsection{VLA Data}
%observations done on 2018/03/23
% 2018-03-23T15:14:20.999999
% On source time from obs prep tool: 1hr 07 min
We performed observations with the VLA on 2018 March 23, 14 days after the detection of the FRB. The VLA was in A-configuration, and the observing project code was VLA/18A-462. The central observing frequency was 2999 MHz, with a total bandwidth of 2048 MHz divided into 1024, 2\,MHz-wide contiguous frequency channels, and two polarizations.  Our pointing center was directed at position R.A.$=21^{\rm h}24{^{\rm m}}19{^{\rm s}}.15$, Decl.=$-33\degree56'10''$ (J2000), which is the pointing center reported by \citet{oslowski2019}. 
We used a 2\,s sampling interval. The standard primary calibrator 3C48 was used for flux density and bandpass calibration, and source J2109-4110 was used as a phase calibrator.  We obtained a total on-source time of 1.15\,hr. 

We calibrated the data using the standard VLA calibration pipeline, followed by manual flagging and imaging with the {\sc casa} software package\footnote{\url{https://casa.nrao.edu/}}. We interactively deconvolved the images using the {\sc tclean} task, with natural weighting to maximize image sensitivity. We obtained an RMS of 5.7\,$\mu$Jy\,beam$^{-1}$ in the central regions of our final image. The synthesized beam major and minor axes were 1.86$\arcsec$ and 0.67$\arcsec$ respectively, and the beam position angle was -8.6$\degree$.

%employing multi-frequency synthesis and...} \fixme{add other non-standard imaging parameters that you used here. Report the weighting algorithm (briggs, natural, uniform) that you used and what robustness parameter, if relevant. Note how much of the primary beam you analyzed and if you performed primary beam correction for the image you show in the figures. Then report: the RMS of the final image at the center of the image, and the synthesized beam major/minor axes and position angle.}

%repeats \ch{The data were taken with 33 spectral windows.} 
%\ch{Spectral windows 12, 15, 16, and 17 were excised due to major \ch{radio-frequency interference}. Spectral windows 29-33 showed gain noise and were also excised. }

\subsection{Optical Imaging Data}
% [Nico to summarize parameters of any Gemini imaging.]
% Gemini imaging griz, dates, etc.
%(program GS-2018B-Q-133 for u-band, not used)
%(program GS-2018A-Q-205 for g,r,i,z)
We observed the field towards \src\ using the Gemini Multi-Object Spectrograph \citep[GMOS;][]{gmos, gmos-s} on Gemini-South. We obtained a set of $1\times 300$\,s images in the $i$ and $z$ bands on UT 2018 March 28 (19 days after the FRB) and a set of $3\times 300$\,s and $4\times 300$\,s images in the $g$ and $r$-bands, respectively, on UT 2018 April 19 (41 days after the FRB). We reduced and co-added these images with {\sc Pyraf}\footnote{\url{https://www.gemini.edu/sciops/data-and-results/processing-software}} using standard procedures. In the $z$ band, before combining, we remove the fringe pattern using a fringe map provided by the Gemini Observatory.
The astrometric solution for the images was computed using 16 stars identified in our $r$-band image, which were also  present in the GAIA DR2 catalog. Comparing the positions of these stars, we obtained a root mean square uncertainty of $0.038\arcsec$ and $0.041\arcsec$ in R.A. and Decl. respectively.
We performed optical photometry for the sources within the uncertainty region to the \src\ (see Section~\ref{sec:loc} for details of localization region) using SExtractor package \citep{Bertin1996} with the $r$-band image, which  was the most sensitive among all the images, adopted as the reference for source detection. We measured the galaxy magnitudes in all bands using flux\_auto with an aperture of 2.5 times the Kron radius, which includes $>$96\% of the total flux of the galaxy \citep{Kron1980}. We corrected these magnitudes to the total flux by measuring the growth curve of isolated stars out to a radius of 10\arcsec. As visible in Table~\ref{tab:src_loc}, the faintest detected galaxy has a $r$-band magnitude of 24.27, which can provide an estimation of the achieved completeness. 

%Our $r$-band image was the most sensitive among all the images, and we reached a completeness of 24.27~mag in this band. We used this $r$-band image for the main analysis reported in this paper. 

We estimated photometric redshifts for these galaxies using the {\sc Eazy} software \citep{Brammer2008} and report the 95\% c.l. values; the Galactic extinction is low in the field direction \citep{schlafly2011}, and thus we did not correct it. 

%We found $20$ sources within the uncertainty region of \src, and their optical magnitudes and estimated photometric redshifts are summarized in Table~\ref{tab:src_loc}. 

We performed an astrometric comparison of the VLA radio and optical images to check and correct any systematic offset between the two. We identified three sources lying outside the localization region of the FRB, which were clearly detected in both the images. These sources were above a significance of 7$\sigma$ in radio and above the optical magnitude limit.
Details of these sources, along with their respective offsets, are given in Table~\ref{tab:astrometry}. 
We calculated an average offset of 0.132$\arcsec$ in RA and 1.7826$\arcsec$ in DEC between the two images. This offset was corrected for by shifting the radio image. We recognized that the offset in declination was far greater than the phase-referencing astrometric standard typically reached with the VLA; we determined (in private communication with L. Sjouwerman of NRAO) that this offset was likely due to a combination of the relatively large offset between phase calibrator and target ($\sim$8 degrees) and the near-horizon observation that the low declination required of the VLA.

\begin{deluxetable*}{ccccccc}
\tablenum{1}
\tablecaption{Astrometric Comparison \label{tab:astrometry}}
\tablewidth{\linewidth}
\tablehead{
\colhead{Sr. No} & \multicolumn{2}{c}{Radio (J2000)} & \multicolumn{2}{c}{Optical (J2000)} & \multicolumn{2}{c}{Offset ($\arcsec$)}\\
\colhead{} & \colhead{RA} & \colhead{DEC} & \colhead{RA} & \colhead{DEC} & \colhead{RA} & \colhead{DEC}}
\startdata
1 & 21:24:7.0598 $\pm$ 0.0029 & -33:57:28.5982 $\pm$ 0.1876 & 21:24:07.046 & -33:57:30.58 & 0.207 & 1.9818 \\
2 & 21:24:11.8076 $\pm$ 0.0038 & -33:55:34.2159 $\pm$ 0.1020 & 21:24:11.802 & -33:55:35.80 & 0.084 & 1.5841 \\
3 & 21:24:20.418 $\pm$ 0.016 & -33:54:26.528 $\pm$ 0.193 & 21:24:20.411 & -33:54:28.31 & 0.105 & 1.782 \\
% 1 & 21:24:7.061 & -33:57:28.74 & 21:24:07.046 & -33:57:30.58 & 0.225 & 1.84 \\
% 2 & 21:24:11.814 & -33:55:34.22 & 21:24:11.802 & -33:55:35.80 & 0.18 & 1.58 \\
% 3 & 21:24:20.444 & -33:54:26.55 & 21:24:20.411 & -33:54:28.31 & 0.495 & 1.76 \\
% 1 & 21:24:7.048 & -33:57:30.51 & 21:24:07.046 & -33:57:30.58 & 0.002 & -0.07 \\
% 2 & 21:24:11.801 & -33:55:35.66 & 21:24:11.802 & -33:55:35.80 & 0.001 & -0.14 \\
% 3 & 21:24:20.431 & -33:54:28.31 & 21:24:20.411 & -33:54:28.31 & 0.02 & 0 \\
\enddata
\end{deluxetable*}

%[Xavier to summarize any Keck imaging.]

% \subsection{Optical Spectroscopy (?)}
% Xavier to write, if we do any actual spectroscopy.

\subsection{ATCA Data}
% The below is copied from emails from Jane Kaczmarek Sun, Mar 31.
The brightness of \src\ led \citet{oslowski2019} to search for the signature of \hi\ absorption in the spectrum of this burst. They reported the ``most prominent'' absorption feature (2.8$\sigma$; private correspondence) at 1386\,MHz, implying a source redshift of $z\,=\,0.025$. The redshift of this feature is consistent with the redshift upper limit for this FRB (see Section~\ref{sec:dm}). Therefore, we collected data with the Australia Telescope Compact Array (ATCA) to search for \hi\ signatures of host galaxies. 

We took the data on 2018 July 19, 20, and 21 (132-134 days after the FRB) with the telescope in the hybrid 75\,m array and recorded in both continuum and high-frequency resolution ``zoom'' modes. Continuum data has a center frequency of 2100 MHz with 2048, 1-MHz channels. We recorded a ``zoom'' band, centered on 1386 MHz with a velocity resolution of 0.11 km\,s$^{-1}$ and spanning the velocity range 1189.4 km\,s$^{-1}$.

% \fixme{The telescope was pointed at the location of the two optical galaxies, as we did not know the redshifts of the optically identified galaxies within the field of view, we could not point to the anticipated \hi\ locations. 

The telescope was pointed at R.A.$=21^{\rm h}24{^{\rm m}}20{^{\rm s}}.27$, Decl.=$-33\degree56'35''.10$ (J2000) and 
R.A.$=21^{\rm h}24{^{\rm m}}16{^{\rm s}}.65$, Decl.=$-33\degree55'57''.20$ (J2000),
% 21:24:20.27, -33:56:35.10 and 21:24:16.65, -33:55:57.20, 
which corresponds to the positions of the two optically-defined galaxies, m402-023014 and m402-023436 \citep{mrss}. We did not know the redshifts of the optically identified galaxies within the field of view. Therefore, we could not point to the anticipated \hi\ locations. As these pointings are within the same ATCA primary beam, we treat the pointings as a mosaic and combine the datasets with a total on-source time of 7.25 hours. We used standard calibrator PKS 1934$-$638 for bandpass and absolute flux calibration. Observations of PKS 2149$-$306 were taken every 40 minutes for phase and gain calibration.
%JK will find reference for these galaxy classifications$

We reduced the data using the {\sc miriad} \citep{miriad} software package using standard routines. We mainly used the automated task {\sc pgflag} for flagging of the data, with minor manual flagging using tasks {\sc blflag} and {\sc uvflag}. As the two pointing centers were within half the width of the primary beam, we mosaiced the two pointings together and made a naturally-weighted total intensity map using the entire 2\,GHz continuum bandwidth with a synthesized beam of $20\times15\arcsec$ and an RMS of 0.1\,mJy\,beam$^{-1}$. We made the images using the compact antennas only, and excluded antenna 6 as it is located 6\,km from the main array.

\section{The DM-based redshift of \src}\label{sec:dm}
The observed DM$_{\text{FRB}}$ for \src\ as quoted by \citet{oslowski2019} is $263.42\pm0.01$\,\dmunits. We use the {\sc ``frbs''} library \citep[][]{frbs} presented first in \citet{prochaska2019} to determine local contributions to this observed DM based on its position: those from the Milky Way's interstellar medium (DM$_{\text{MW}}$) and from our galaxy's halo (DM$_{\rm Halo}$). We estimated DM$_{\text{MW}}$ using two electron density models of the Galaxy: NE2001 \citep{cordes2002} and YMW16 \citep{yao2017}. 
%\mbox{(DM$_{\text{MW}}=46\,$\dmunits\ and 30\dmunits} using NE2001 \citep{cordes2002} and YMW16 \citep{yao2017} models) and from our galaxy's halo \mbox{(DM$_{\rm halo}=63\,$\dmunits).} 
For this analysis, we also assume a conservatively small host-galaxy DM contribution of DM$_{\text{host}}=50\,$\dmunits, which serves to account for a fairly standard contribution from a host galaxy's ISM and halo \citep{prochaska2019}. This value of DM$_{\text{host}}$ is also consistent with the empirically derived 95\% confidence interval on DM$_{\text{host}}$ obtained by \citet[][]{Macquart2020}.\footnote{Note that Macquart et al.\ determined a galaxy-rest-frame average host DM$_{\rm rest}$ of 100\,\dmunits.
However, this value is closer to our estimate here when considering the contribution to the observed DM will scale by 
${\rm DM_{\rm rest}}/(1+z)$.}

%based on the analysis of localized FRBs in \citet{Macquart2020}.}%can be estimated at the location of \src\ from electron density models of the galaxy (\eg\ NE2001 \citealt{cordes2002} and YMW16 \citealt{yao2017}). These give similar 
%We use the estimate by  \citep{prochaska2019} and then assume the Halo contribution to be 65 $\pm$ 15 \dmunits (similar to \citep{pol2019}). 
We then used the dispersion measure of \src\ to estimate the likely redshift of an FRB host galaxy by subtracting the various DM contributions listed above. These estimates imply a likely range on the DM contribution from intergalactic medium (DM$_{\text{cosmic}}$) to be  104--120\,\dmunits\ (see Table~\ref{tab:dm_igm_redshifts}). These numbers may, of course decrease if the host DM contribution is significantly larger than what we have assumed.
%We consider these values as maximum likely redshift values because they disinclude the likely greater contribution from a host (Note, the mean host value fitted by \citet{Macquart2020} was 100\,\dmunits), and because we do not include estimates here of intervening galaxy haloes that are likely to add contributions to DM$_{\rm FRB}$

\begin{figure}
    \centering
    \includegraphics[height=0.4\textwidth]{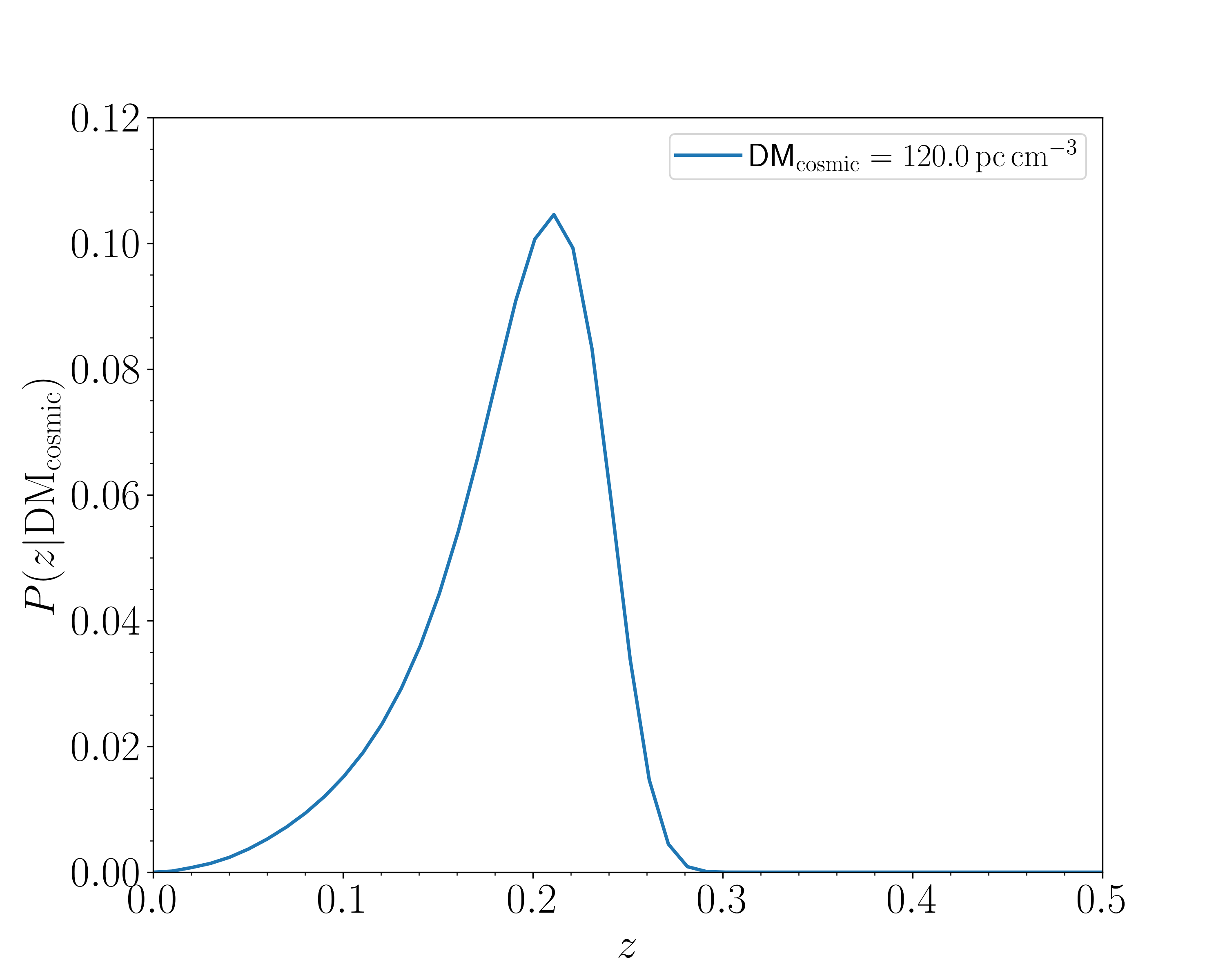}
    \caption{Probability density function $P(z|\text{DM}_{\text{cosmic}})$ for \dmcosmic\ = 120\dmunits. 
    }
    \label{fig:PDM_z}
\end{figure}

%JXP -- New text
From the \dmcosmic\ value above, we may estimate a firm upper limit to the
FRB redshift and also a best estimate.  For the latter, we adopt the
mean Macquart relation\footnote{\url{https://github.com/FRBs/FRB}}
with the cosmological parameters from \citep{planck2015}.
This yields $z=0.13$.
For the upper limit, we adopt the probability distribution function
for \dmcosmic\ from \cite{Macquart2020} and assume a uniform prior 
in redshift to assess $P(z|DM)$ as depicted in Figure~\ref{fig:PDM_z}.
From our upper limit to \dmcosmic\ we set a conservative upper
limit $z<0.32$. Wherever necessary, we use the above two redshift values to estimate the luminosities of relevant sources. It is possible that the FRB (and its host galaxy) is much closer than the DM-based redshift limit derived above, with a large fraction of the dispersion measure being contributed by plasma local to the FRB.

\begin{deluxetable}{ccc}
\tablenum{2}
\tablecaption{Estimate of DM$_{\text{cosmic}}$ of \src \label{tab:dm_igm_redshifts}}
\tablewidth{\linewidth}
\tablehead{
\colhead{DM part} & \colhead{NE2001} & \colhead{YMW16}
% \colhead{} & \colhead{DM_{\text{IGM}} 1} & \colhead{DM_{\text{IGM}} 2} & \colhead{DM_{\text{IGM}} 1} & \colhead{DM_{\text{IGM}} 2}
} 
\startdata
DM$_{\text{FRB}}$ & 263 & 263 \\
DM$_{\text{MW}}$ & 46 & 30 \\
DM$_{\text{Halo}}$ & 63 &  63 \\
DM$_{\text{host}}$ & 50 & 50  \\
DM$_{\text{cosmic}}$ & 104 & 120  \\
\hline
% z$_1$ & 0.09 & 0.11 \\
% % z$_2$ & 0.59 & 0.6 \\
% z$_2$ & 0.48$_{-0.09}^{+0.06}$ & 0.51$_{-0.10}^{+0.06}$ \\
\enddata
\end{deluxetable}

% \remove{
% There appears to be a discrepancy in the redshifts estimated by the two approaches. We briefly discuss the reason for this here (see Section 3.3 of \citet{pol2019} for a detailed discussion). The underlying distribution of DM$_{\text{IGM}}$ was found to be highly skewed by \citet{pol2019}, especially at low redshifts. Therefore, approaches like \citet{ioka2003, inoue2004} which use analytical expressions representing the average behavior of IGM, overestimates DM$_{\text{IGM}}$ at low redshifts. \citet{pol2019} on the other hand, used $N$-body dark matter simulations to track the dark matter particle number density over a large range of redshifts. They used this to obtain the free electron density, and calculate a DM as a function of redshift from it. They are therefore able to account for the intrinsic skewness in DM$_{\text{IGM}}$. This is supported by the empirical scatter in the DM vs.\ $z$ relation reported by  \citet[][]{Macquart2020}. The empirical scatter in this ``Macquart Relation'' demonstrates contributions from both host redshift and intrinsic variations in DM$_{\rm IGM}$ along various lines of sight.
% We thus consider these fiducial values as a nominal range for the FRB redshift because they encompass the range of expected natural variations in the IGM and in the intersection of intervening halos \citep[\eg][]{prochaska2019}.
% }

%Location of \src 21:24:43, -33:58:44. 

\section{The localization of \src}\label{sec:loc}

% \begin{figure*}
%     \centering
%     \includegraphics[width=\textwidth]{figures/beam_fit.png}
%     \caption{Localization center (black point) in the response of each beam of the Parkes multibeam detector, as seen for the lower third of our observing band.
%     % Beam 6 is excluded
%     The axes denote focal plane coordinates \verify{relative to the fitted position of the FRB}. 
% }
%     \label{fig:beamloc}
% \end{figure*}

% \fixme{{\bf Vikram to review/revise this section!}}

%We need a description of how \src\ was localized and how good the localization region was, with associated beam figures.}
If we know the side-lobe structure of the Multibeam Receiver, we can use information about the beam-dependent signal to noise (S/N) of a detected source to localize an FRB to a region smaller than the oft-quoted half-power beamwidth of the telescope (for Parkes at our frequency, this leads to a typical localization region of approximately 150~arcmin$^2$). To improve the localization of \src\, we performed a procedure similar to that of \citet{ravi2016}, in which the high-significance detection of FRB\,150807 in multiple beams allowed us to constrain its position to a 9~arcmin$^2$ region. \src\ was detected 
%\fixme{at a S/N of $>$15} 
in 7 of 13 beams in the Parkes multi-beam receiver. It was saturated in beam 1, supplying only a lower limit to the actual intensity. Treating beam 1 as a lower limit, we followed the procedure as described in \citet{ravi2016}. 
The resulting position following their prescription allowed localization of \src\ to approximately $\sim 2' \times 2'$, as shown in Figure~\ref{fig:err_region}. 

%with \fixme{some\#} confidence.

%``This provides a few arcmin^2 localization (see attached). At the moment, there’s no need to do anything fancier.''
%The central position of the region is 21:24:19, -33:56:12 (J2000). Stefan / Ryan - it would be good to see whether this position makes sense given your idea of where beams 1, 2 and 7 were at the time of the burst. 

\section{Results}\label{sec:results}
% \input{tab_sources.tex}
% \begin{itemize}
%     % \item {\bf Table of sources in the localization region:} radio/optical position, general optical color or maybe just u g r i magnitudes, photo-z, VLA integrated and peak flux (or noise limit) at that loc, probability of association Bloom (no DM) or Eftekhari method (no DM/using DM) or maybe one column for each.
%     % \item {\bf Figure:} Localization region: overlay of optical and radio.
%     \item Notes on individual interesting sources\\
%     --$>$ Ones detected in both radio and optical in localization region (including image cutout).\\
%     --$>$ Z-shaped thing (including image cutout overlay)
% \end{itemize}

\subsection{Objects within Localization Region}
\begin{figure*}
    \centering
    \includegraphics[height=1\textwidth, angle=-90,trim=15mm 15mm 10mm 15mm, clip]{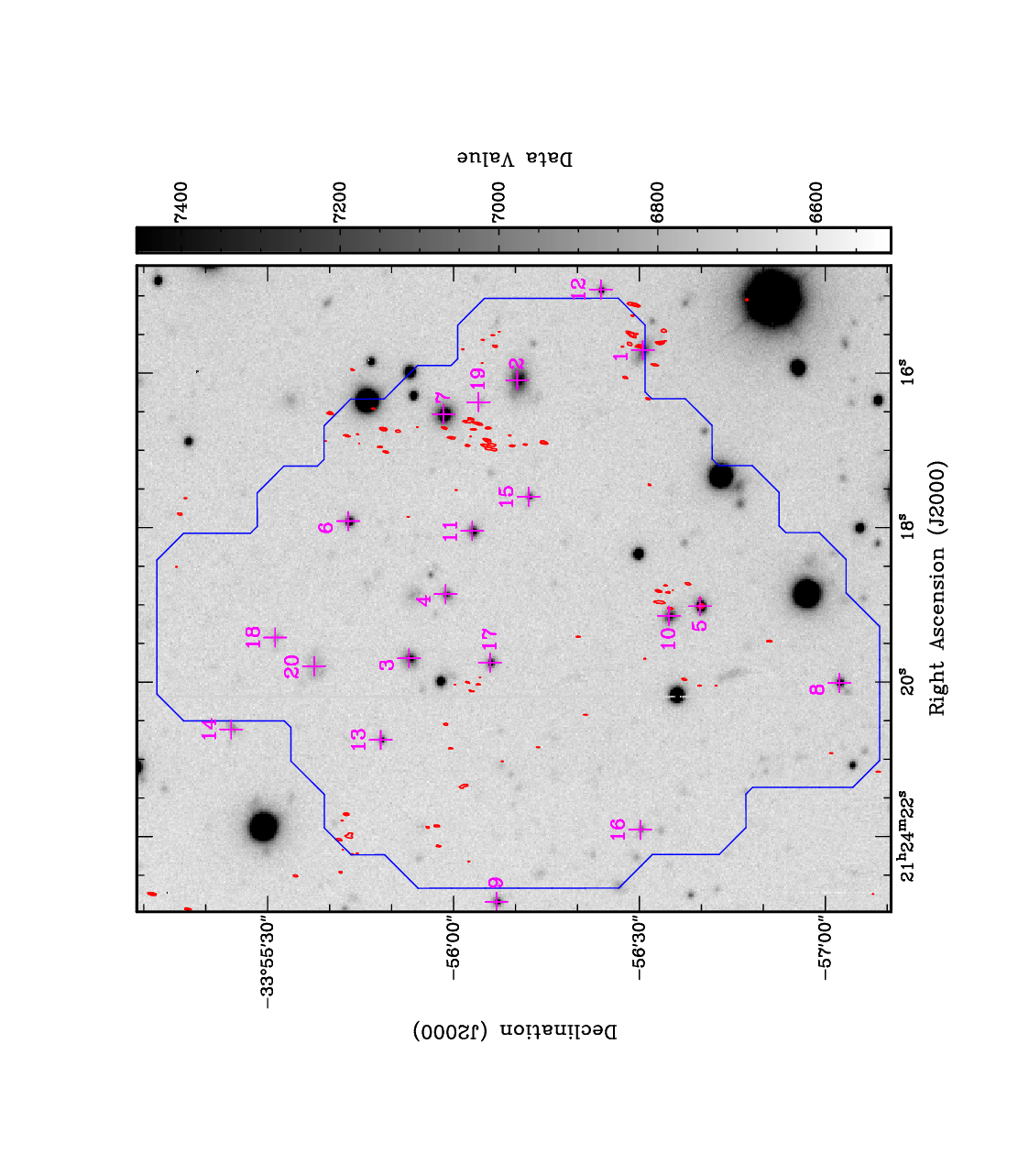}
    \caption{Overlay of radio contours from the VLA (red) on the greyscale GMOS r-band image. The contour levels are 3, 3.5, 4 times 5.7\,$\mu$Jy\,beam$^{-1}$, which is the RMS noise of the off-source central regions of the image. The boundary of the localization region discussed in Section \ref{sec:loc} is shown by a blue contour. %\fixme{The dark vertical feature at a right ascension of 21$^{\rm h}$24$^{\rm m}$23$^{\rm s}$ \verify{is caused by a CCD chip boundary.}}
    The synthesized radio beam size of this image is quoted in Section~\ref{sec:data}. Magenta crosses and numbers represent the positions and source numbers of galaxies in Table~\ref{tab:src_loc}. The length of the crosses is arbitrary and does not represent the positional uncertainty. Stellar objects in the localization region above our magnitude limits are reported in Table~\ref{tab:src_loc}.}
    \label{fig:err_region}
\end{figure*}

We display our radio and optical imaging of the full localization error region in Figure~\ref{fig:err_region}. There were no prominent ($\geq5\sigma$) detections in the VLA radio data within or near the localization error region. 
%There were also no prominent detections in the area of sky $\pm1'$ from the center of the localization error region ($>$4 times the local RMS noise in the image).

Above an $r$-band magnitude limit of 24.27, 20 galaxies were detected in the Gemini GMOS data within the error region. Basic measurements for these galaxies are provided in Table~\ref{tab:src_loc}. Out of these 20 sources, 14 sources have redshift limits consistent with the range of redshifts we estimated for \src\ in \S\ref{sec:dm}. The last column in Table~\ref{tab:src_loc}  gives the association probability of the optical galaxy detections, as discussed later in Section~\ref{sec:assoc}.

We did not detect any steady nor transient radio source within the localization error region, down to a $5\sigma$ limit of $150\mu$Jy\,beam$^{-1}$, in our three-epoch ATCA data. There was no detection of \hi\ counterparts within the localization region, at the implied redshift of the spectral feature reported by \citet{oslowski2019}.

%Following the standard calibration and imaging procedure, along with manual flagging, the achieved RMS value of the VLA radio image was 5.7$\mu$Jy. The radio image overlayed\footnote{In this paper, radio image refers to the image from the VLA data. Images from other radio telescopes will be explicitly stated.} on the optical image is shown in Figure~\ref{fig:err_region}. 

% \begin{enumerate}
%     \item Figure: Radio over optical GMOS r band image, of error region. Radio contours: 3,4,5 sigma
%     \item Coinciding images show that there are two optical sources within the error region. 
%     \item As can be seen in table~\ref{tab:src_loc}, \textbf{blah} of the sources have a photometric redshift similar to that predicted for \src\, and therefore we did a deeper analysis of these sources.
% \end{enumerate}

\begin{deluxetable*}{cccccccccccc}
\tablenum{3}
\tablecaption{Optical sources detected within the localization region. Phot. Redshift is the photometric redshift of the source. VLA $F_{\rm peak}$ is the peak flux in the VLA radio image at the source location. Chance Assoc. Prob. (EB17) is the chance association probability of the source with \src. \label{tab:src_loc}}
\tablewidth{\linewidth}
\tablehead{
\colhead{Source}  & \multicolumn{2}{c}{J2000} & \multicolumn{4}{c}{Photometric magnitudes (AB)} & \colhead{Phot.} & \colhead{VLA $F_{\rm peak}$} & \colhead{Chance Assoc.} \\ 
\colhead{No.} & \colhead{R.A.} & \colhead{Dec.} & \colhead{$g$} & \colhead{$r$} & \colhead{$i$} & \colhead{$z$} & \colhead{Redshift\tablenotemark{a}} & \colhead{($\mu$Jy/beam)\tablenotemark{b}} & \colhead{Prob. (EB17)\tablenotemark{c}}} 
% \colhead{Sr. No} & \multicolumn{2}{c}{Radio (J2000)} & \multicolumn{2}{c}{Optical (J2000)} & \multicolumn{2}{c}{Offset ($\arcsec$)\\
% \colhead{} & \colhead{RA} & \colhead{DEC} & \colhead{RA} & \colhead{DEC} & \colhead{RA} & \colhead{DEC}}
\startdata
1&21:24:15.701&-33:56:30.55&$22.36\pm0.07$&$21.60\pm 0.04$&$20.46\pm 0.08$&$20.87\pm 0.05$&$0.16^{+0.53}_{-0.07}$&24.28&0.99\\
2&21:24:16.090&-33:56:10.29&$21.12\pm0.06$&$20.47\pm 0.04$&$19.78\pm 0.08$&$19.61\pm 0.04$&$0.31^{+0.41}_{-0.25}$&$<$19.5&0.76\\
3&21:24:19.693&-33:55:52.82&$22.22\pm0.06$&$21.60\pm 0.04$&$21.05\pm 0.08$&$21.03\pm 0.06$&$0.42^{+0.26}_{-0.33}$&$<$16.8&0.99\\
4&21:24:18.860&-33:55:58.73&$23.10\pm0.07$&$22.43\pm 0.04$&$21.99\pm 0.09$&--&$0.47^{+0.66}_{-0.43}$&$<$16.5&1.0\\
5&21:24:19.018&-33:56:39.80&$22.00\pm0.06$&$21.25\pm 0.04$&$20.46\pm 0.08$&$20.23\pm 0.03$&$0.52^{+0.25}_{-0.44}$&22.1&0.96\\
6&21:24:17.916&-33:55:43.00&$22.28\pm0.06$&$21.81\pm 0.04$&$21.19\pm 0.08$&$21.09\pm 0.05$&$0.53^{+0.24}_{-0.45}$&$<$17.1&1.0\\
7&21:24:16.529&-33:55:58.40&$21.71\pm0.06$&$20.38\pm 0.04$&$19.36\pm 0.07$&$19.20\pm 0.03$&$0.55^{+0.11}_{-0.34}$&$<$19.2&0.73\\
8&21:24:20.012&-33:57:02.30&$22.85\pm0.07$&$21.93\pm 0.04$&$21.04\pm 0.08$&$20.93\pm 0.05$&$0.56^{+0.17}_{-0.44}$&$<$15.72&1.0\\
9&21:24:22.844&-33:56:06.92&$22.56\pm0.07$&$22.38\pm 0.04$&$22.19\pm 0.10$&--&$0.62^{+0.64}_{-0.57}$&$<$15.93&1.0\\
10&21:24:19.142&-33:56:34.77&$22.44\pm0.07$&$21.71\pm 0.04$&$20.84\pm 0.08$&$20.85\pm 0.06$&$0.65^{+0.09}_{-0.57}$&$<$17.7&0.99\\
11&21:24:18.042&-33:56:03.02&$22.32\pm0.06$&$21.98\pm 0.04$&$21.26\pm 0.08$&$21.54\pm 0.08$&$0.66^{+0.09}_{-0.58}$&$<$17.85&1.0\\
12&21:24:14.916&-33:56:23.78&$23.80\pm0.08$&$22.66\pm 0.04$&$21.58\pm 0.08$&$21.26\pm 0.06$&$0.66^{+0.14}_{-0.15}$&$<$18.9&1.0\\
13&21:24:20.748&-33:55:48.25&$24.12\pm0.09$&$22.94\pm 0.04$&$21.37\pm 0.08$&$21.09\pm 0.05$&$0.70^{+0.18}_{-0.09}$&$<$17.4&1.0\\
14&21:24:20.612&-33:55:24.12&$23.97\pm0.09$&$23.13\pm 0.05$&$22.43\pm 0.11$&--&$0.71^{+0.70}_{-0.57}$&$<$18.0&1.0\\
15&21:24:17.600&-33:56:12.12&$23.80\pm0.08$&$22.80\pm 0.04$&$21.39\pm 0.08$&$20.98\pm 0.05$&$0.76^{+0.17}_{-0.13}$&$<$18.6&1.0\\
16&21:24:21.913&-33:56:30.13&$23.85\pm0.08$&$23.38\pm 0.05$&$22.79\pm 0.11$&--&$0.82^{+0.70}_{-0.67}$&$<$15.9&1.0\\
17&21:24:19.748&-33:56:05.88&$22.43\pm0.06$&$22.34\pm 0.04$&$21.32\pm 0.08$&$20.97\pm 0.06$&$0.87^{+0.40}_{-0.12}$&$<$16.8&1.0\\
18&21:24:19.423&-33:55:31.22&$23.43\pm0.08$&$23.43\pm 0.06$&$23.04\pm 0.12$&--&$0.99^{+0.61}_{-0.80}$&$<$17.7&1.0\\
19&21:24:16.375&-33:56:04.05&$24.66\pm0.11$&$24.27\pm 0.09$&$23.43\pm 0.14$&--&$1.08^{+1.02}_{-0.57}$&$<$19.2&1.0\\
20&21:24:19.793&-33:55:37.49&$22.46\pm0.07$&$23.89\pm 0.09$&$21.13\pm 0.08$&$21.14\pm 0.06$&$1.45^{+0.16}_{-0.10}$&$<$17.4&1.0\\
\enddata
\tablenotetext{a}{Photometric redshifts correspond to the $z_{\rm peak}$ parameter from EAZY, and the uncertainties correspond to the $95\%$ c.l. interval.}
\tablenotetext{b}{For sources with intensity less than the local $3\sigma$ image RMS, we have reported the $3\sigma$ RMS. The intensities of source 1 and 5 are more than $3\sigma$, but 
the signal-to-noise are low ($<6$), so we have reported flux at the peak pixel.}
\tablenotetext{c}{Probability was calculated using $r$-band magnitudes after correcting for extinction, A$_\lambda=0.179$ \citep{schlafly2011}.}
\end{deluxetable*}

\subsection{Coincident Radio/Optical Detections}\label{sec:coincident}
\begin{figure*}
    \includegraphics[angle=-90,width=\textwidth, trim=4.5cm 0cm 4.25cm 1.5cm, clip]{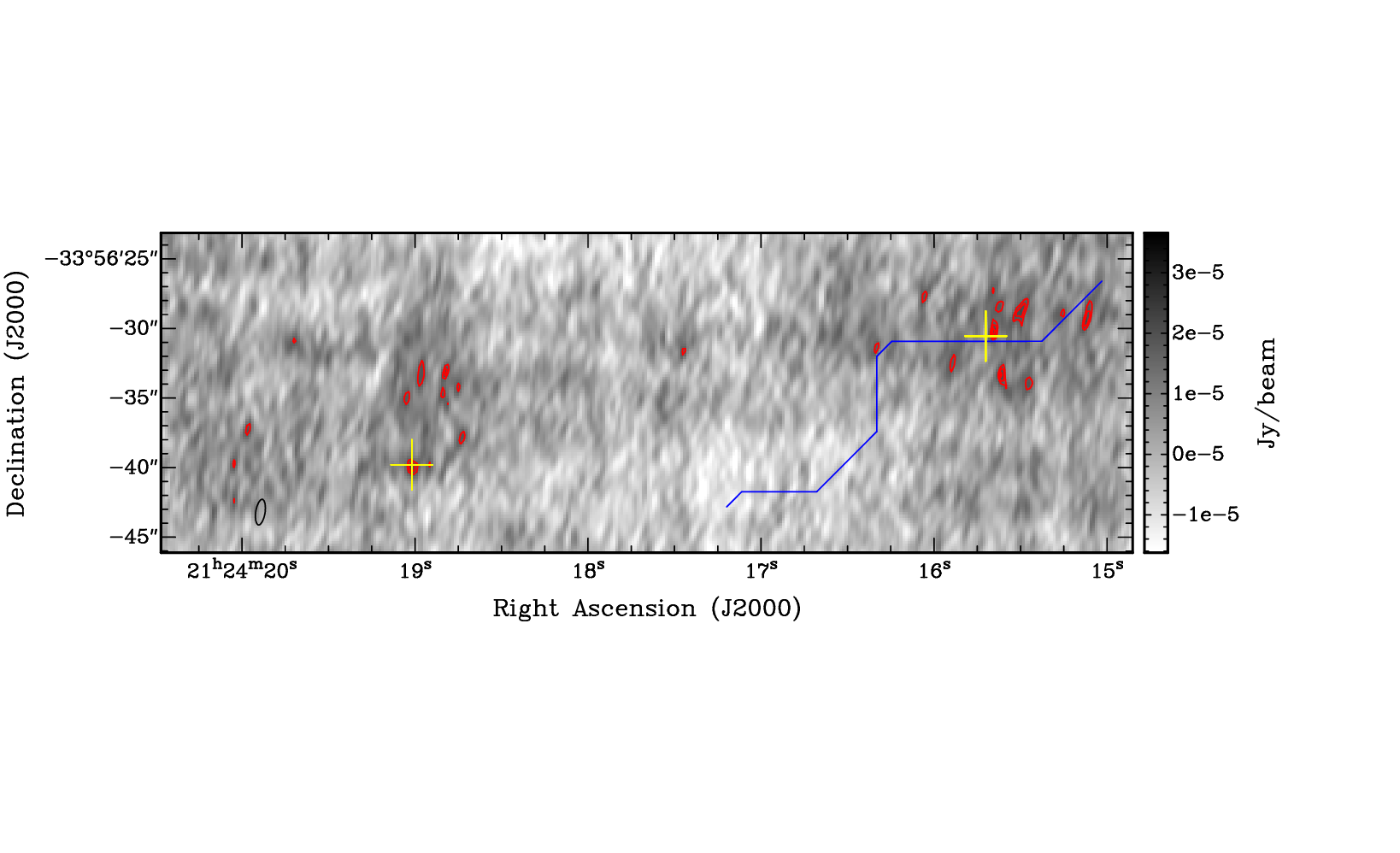}
  \caption{Sources within the localization region with a radio - optical overlap. The background grayscale and red contours are from VLA radio data. The boundary of the localization region is shown with a blue contour. The best fit optical position of the galaxies is shown with yellow crosses. The radio contours levels are placed at 5.7\,$\mu$Jy\,beam$^{-1}$ times [3, 3.5, 4]. VLA beam is shown as a black ellipse in the bottom left.}\label{fig:interesting_src}
\end{figure*}

The brightest radio feature within the localization error region had a S/N of only 4.3 in our VLA imaging (corresponding to a peak flux of approximately 24.3\,$\mu$Jy\,beam$^{-1}$). Thus, while we did not detect any prominent radio sources in this field, there were two marginal radio detections (the ${\rm S/N=4.3}$ event and one at ${\rm S/N=3.9}$) that were coincident with our optical galaxy identifications; these fluxes are reported in Table~\ref{tab:src_loc} for sources 1 and 5.
%We detected two objects within the localization error region, with a $>3\sigma$ detection in VLA radio image and near an optical galaxy. 
An enlarged region showing these targets is displayed in Fig~\ref{fig:interesting_src}; it is clear from this figure that while these may be genuine detections, residual low-level side-lobe features in the image may be artificially boosting the flux at these locations. These detections are discussed further in Section~\ref{sec:int_obj}. %The photometric redshifts of the optical sources near them and the chance association probability of these sources with \src\ are given in Table~\ref{tab:src_loc}. 

%\fixme{None of these sources correspond to any archival source.} 

\subsection{Serendipitous Detection of a Winged Field Object}
\begin{figure}
    \centering
    \includegraphics[trim={2.3cm 5.7cm 1.0cm 5.5cm},clip, angle=-90, width=1.0\columnwidth]{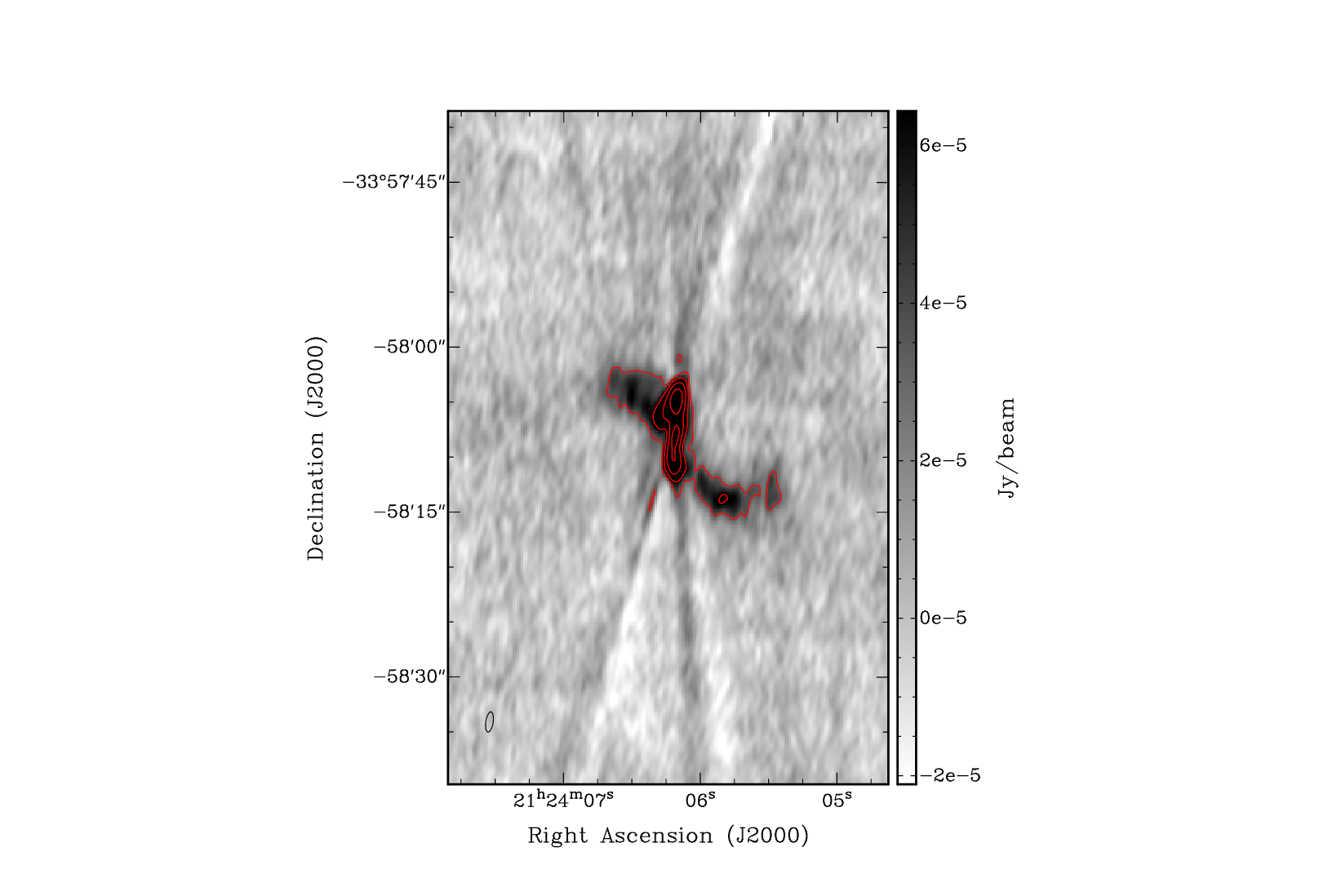}
    \caption{Radio image of the serendipitous (unrelated) source detection in our field. The background image and red contours are from VLA radio data. The radio contours levels are placed at 5.7\,$\mu$Jy\,beam$^{-1}$ times factors of [6, 12, 24, 48]. Note that the straight diagonal lines are due to side lobes and are an artifact of the imaging process.}
    %Note the high significance of the two cores at the center.}
    \label{fig:z_source}
\end{figure}

In our imaged field, but far from the localization region of \src, we detected a complex radio source (hereafter referred to as \xsrc). We show this object in 
Fig~\ref{fig:z_source}. A compact optical galaxy is co-located at the nexus of the complex structures and is presumably the host. The position of the central optical source in the Gemini $r$-band image is 
%R.A.$=21^{\rm h}24{^{\rm m}}6{^{\rm s}}.171$, Decl.=$-33\degree58'8''.32$ (J2000) [\fixme{casa fit: 
J2000 R.A.$=21^{\rm h}24{^{\rm m}}6{^{\rm s}}.171(2)$, Decl.=$-33\degree58'8''.3(2)$. As this presumed host was at the edge of the optical image, the magnitudes in different optical bands were not reliable enough to determine its photometric redshift. %\verify{A search of the NASA extragalactic database\footnote{http://ned.ipac.caltech.edu/} did not reveal any cataloged source within $2\arcmin$ of \xsrc\ in an archival search.} 
\xsrc\ is very similar in complexity to the X-, S-, and Z-shaped sources studied by \citet{cheung2007, roberts2015, saripalli2018, roberts2018, lal2019, joshi2019}. This appears to be an uncataloged example of such galaxies, as a cross-check of published lists of X-shaped sources \citet{cheung2007, yang2019, proctor2011}, did not include this object. 
%Thus we appear to have newly identified it as a source of this class. 
While the object was detected as a 7.7\,mJy beam$^{-1}$ radio source in the 1.4\,GHz NVSS survey \citep{nvss}, the $\sim$45$''$ NVSS resolution caused the survey to detect this object as a point source. 
%We therefore conclude that \xsrc\ has not been detected previously.

%Such galaxies with multi-axis morphologies are characterized by a small axial ratio with off-axis emission. 
%Such multi-axis morphologies are seen in a small number of radio galaxies with two pairs of collimated radio lobes, and have radio luminosity which is intermediate to FR I and FR II type galaxies, indicating that they might represent a transitional morphology between the two \citep{landt2010}. The origin of X/S/Z-shaped radio galaxies is still unclear, and three main models are: (1) diversion of the backflowing synchrotron plasma of radio lobes upon impacting an asymmetric circum-galactic gaseous halo of the parent early-type galaxy \citep{leahy1984}; (2) spin-flip in the central super massive black hole \citep{zier2001} and (3) ongoing precession of the twin jets in a dual-AGN \citep{lal2004}. 

% joshi have explaied o-dev galaxy by some idea. 

{}

\section{Discussion}\label{sec:discussion}
\subsection{Association Probability of Sources}\label{sec:assoc}
\defcitealias{eftekhari2017}{EB17}
\defcitealias{bloom2002}{B02}
% In text \citet{jd14}. Or in brackets \citep[][hereafter JD14]{jd14}.
We followed three approaches to calculate the probability of chance coincidence of the FRB with the sources in the error region. We followed two procedures described in \citet[][Sec.~2 and 3]{eftekhari2017} to calculate the probability without and with using the estimated redshift of the FRB. The first approach assumes a Poisson distribution of radio sources across the sky and calculates the chance coincidence probability using the number density of galaxies above a limiting $r$-band magnitude. In the second approach, the number density of galaxies at a given redshift is calculated by integrating the optical luminosity functions. We also use the approach of \citet[][Sec.~6.1]{bloom2002}, which used the \textit{r} band magnitude and the expression given by \citet{hogg1997} to calculate the expected number of galaxies within a given radius. This is then used to calculate the corresponding association probability\footnote{Implemented in \url{https://github.com/KshitijAggarwal/casp}}\citep[][]{casp}. Some details of the above methods are given in Appendix~\ref{appendix:assoc}. 

We calculated the chance coincidence probability of all the sources given in Table~\ref{tab:src_loc}, after correcting the $r$-band magnitudes for Galactic extinction \citep{schlafly2011}. We do not report the probabilities calculated using \citet[][]{bloom2002} and redshift approach of \citet[][]{eftekhari2017} in Table~\ref{tab:src_loc}, as they were all $\sim$1. As no bright optical source was seen within the localization region, and because the localization region was large, the chance coincidence probability for all these associations were close to 1. Therefore, we cannot confidently associate \src\ with any of the observed galaxies based on spatial coincidence information alone.

% \subsection{Limits on Persistent Radio Emission}\label{sec:int_obj}
\subsection{Plausible Host Galaxies}\label{sec:int_obj}
Here, we discuss the properties of the detected galaxies within the FRB error region and discuss the implications of their radio/optical properties.

\begin{figure*}
    \centering
    \includegraphics[height=0.5\textwidth, angle=0, trim=0mm 0mm 0mm 0mm,clip]{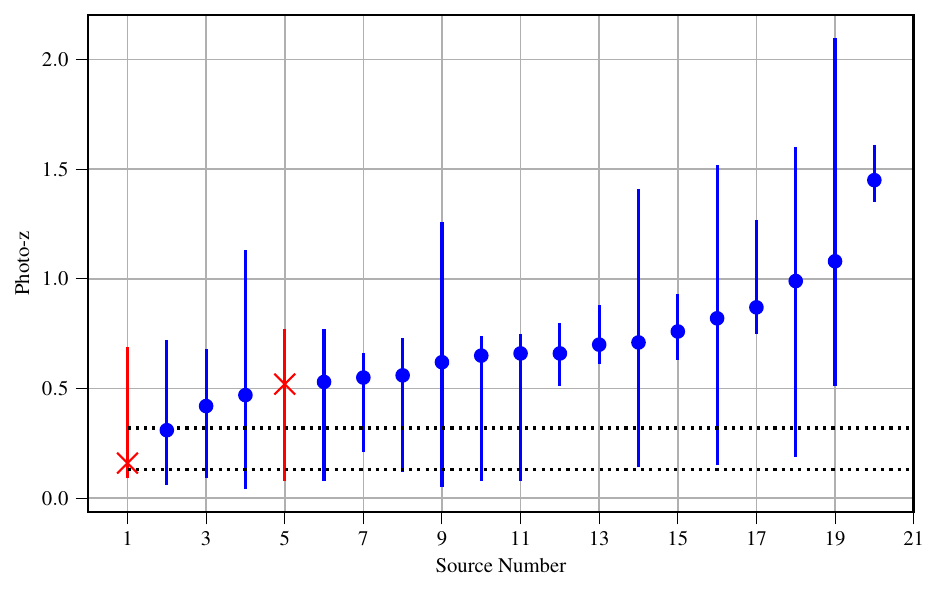}
    \caption{A comparison of the galaxy photometric redshifts with the FRB redshift estimated for two different formulations of the DM-redshift relation (Section \ref{sec:dm}). The blue points with errors represent all galaxies within the localization error region of \src, while the red crosses with errors represent the galaxies with a potential radio association also within the error region. The black dotted line shows the FRB redshift estimates.  It is clear that while some galaxy associations can be ruled out, due to the large errors in our photometric redshift measurements, many galaxies in this sample remain host candidates based on redshift criteria.}
    \label{fig:redshift_fig}
\end{figure*}

\subsubsection{Photometric Redshift Comparison}
In Figure \ref{fig:redshift_fig}, we compare the galaxy photometric redshifts with the range of FRB redshifts previously estimated ($z= 0.13$--0.32). While only two galaxies have most-likely photometric redshift values that lie within our estimated FRB redshift range, the photometric redshift errors are relatively large, thus do not preclude other galaxies from remaining contenders for the FRB host (in addition to dwarf galaxies below the detection limit of our observations, as we discuss in Section \ref{sec:analysisofwhether}). However, five galaxies in this field (source numbers 12, 13, 15, 17, 19, 20 in Table \ref{tab:dm_igm_redshifts}) are unlikely hosts for the FRBs based on redshift information.

%It is clear that while several galaxy associations can be ruled out, due to the large errors in our photometric redshift measurements, many galaxies in this sample remain host candidates based on redshift criteria.

\subsubsection{Using Coincident Radio/Optical Detections}
Thus far, only one FRB---the repeating FRB\,121102---has had a detection of any co-located ``persistent'' (non-bursting) radio emission. Other localized FRBs have reported strict limits on co-located emission, typically limits corresponding to a few tens of $\mu$Jy \citep{bannister2019, prochaska2019, Ravi+2019, marcote20}. Additionally, most progenitor models do predict, to some level, a radio counterpart \citep[][]{frbtheorycat}. It is thus pertinent to search for any galaxies with associated radio emission in this region.

There were only two optical galaxies within the localization region that had a coincident borderline detection in the radio image. These were sources at photometric redshifts of 0.16$^{+0.53}_{-0.07}$ (``source 1'') and 0.52$^{+0.25}_{-0.44}$ (``source 5''). Because the intensity of these sources is not sufficiently confident to perform reliable source fitting (signal-to-noise ratio $<$5), in Table~\ref{tab:src_loc} we report the flux at the peak pixel for each of the two radio components. The fluxes of these sources correspond to luminosities of $1.7\times10^{21}$\,W\,Hz$^{-1}$ and $2.4\times10^{22}$\,W\,Hz$^{-1}$ at their respective photometric redshifts.
%estimates. For source A, the radio position (J2000 RA, Dec 21:24:15.6 -33:56:28.4) is co-located with the optical galaxy but offset from its centroid reported in Table 4; for source B, the radio position (J2000 RA, Dec 21:24:19.0 -33:56:38.1) is consistent with the optical centroid.
The luminosity of source 5 is comparable to that of the persistent radio source (PRS) of FRB\,121102, while that of source 1 is an order of magnitude lower. We do not have a sufficiently strong detection to comment on the precise origin of this emission (star formation, AGN, or other).

The photometric redshifts of these sources are consistent with the redshift ranges we have estimated for \src. But, due to large errors on photometric redshifts, neither link is conclusive for the association. 
%Moreover, the chance coincidence probability of these sources is also large (see Section~\ref{sec:assoc}). 
Further, as previously noted, these two radio sources may feasibly be arising due to diffuse, residual side-lobe structure in our radio image caused by low-level calibration errors and a bright foreground source at J2000 R.A.$=21^{\rm h}24{^{\rm m}}14{^{\rm s}}.749(2)$, Decl.=$-33\degree47'58''.67(8)$. 
%Therefore, it is not possible to confidently associate these sources with \src. 

%\fixme{Some comment about Gemini obs, Given our mag completeness limit, what galaxies did we see/miss, till what redshift.}

% \subsection{Limits on persistent radio emission}\label{sec:persistent}
\subsubsection{Redshift-dependent Radio Luminosity Limit}
\begin{figure*}
    \centering
    \includegraphics[height=0.5\textwidth]{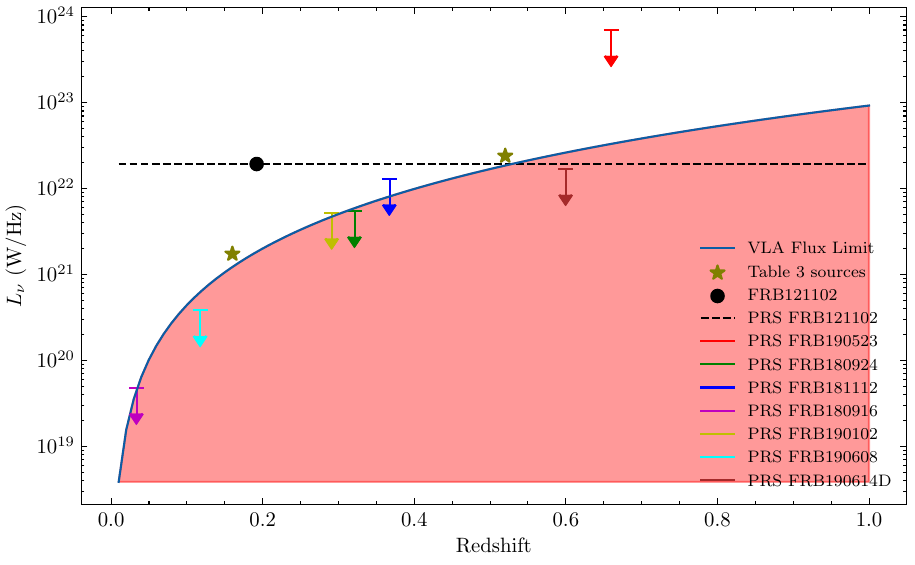}
    \caption{The blue solid line is the luminosity limit (corresponding to VLA flux limit of our observations) with respect to redshift. Black dashed line and point are the luminosity of PRS of FRB\,121102. Upper limits on the luminosity of PRS corresponding to other localized FRBs are shown with downward pointing arrows. Olive stars correspond to the coincident sources discussed in Section~\ref{sec:coincident}. Note that due to large photo-z uncertainty, there is a considerable uncertainty in the luminosity of these coincident sources. Sources within the red shaded region would not be detected with our observations.}
    % Note: had discussion about whether to include error bars on our data points for the table 4 sources. Decided against it but to just leave a note on it in the caption since the error bars are so huge. We will let referee decide if they want to see it.
    \label{fig:lum_plot}
\end{figure*}

The RMS limit of our VLA observations was 5.7\,$\mu$Jy\,beam$^{-1}$ in this field, which corresponds to a $3\sigma$ upper limit of 17.1 $\mu$Jy\,beam$^{-1}$ on the flux of any persistent radio source. This, in turn, corresponds to luminosity limits of $<7.8 \times 10^{20}$\,W\,Hz$^{-1}$ and $<5.8 \times 10^{21}$\,W\,Hz$^{-1}$ at distances corresponding to redshifts of 0.13 and 0.32, %$<5.4 \times 10^{20}$\,W\,Hz$^{-1}$ and $<1.8 \times 10^{22}$\,W\,Hz$^{-1}$ at distances corresponding to redshifts of 0.11 and 0.51, 
respectively (the luminosity limit as a function of redshift is shown in Figure~\ref{fig:lum_plot}). The luminosity limits in this redshift range are lower than the luminosity of the PRS of FRB\,121102 ($< 1.93 \times 10^{22}$\,W\,Hz$^{-1}$). Therefore, if \src\ had a PRS similar to that of FRB\,121102, then we could have detected it out to a redshift of 0.52 (see Fig~\ref{fig:lum_plot}). 

So far, only FRB\,121102 has been co-located with a PRS, with no other localized FRBs having a clearly identified radio counterpart. The redshift-dependent luminosity limits we show in Figure~\ref{fig:lum_plot} are comparable to upper limits from other experiments that have not detected a PRS \citep{marcote20, Macquart2020, law2020, Ravi+2019, bannister2019, prochaska2019}. Given the variety of properties of FRB associations, it is possible that galaxies other than those with detectable radio emission in this field could feasibly remain the \src\ host. 
In addition, as previously indicated, it remains possible that the PRS of FRB\,121102 is unrelated to the FRB or its progenitor, therefore the presence of radio emission might not be specifically indicative of an FRB/host association.

\subsubsection{Analysis of Whether We Detected All Likely Candidate Hosts}\label{sec:analysisofwhether}
Using the magnitude limit of our optical observations, we can estimate the completeness for different galaxy types. Our $r$-band apparent magnitude limit of 24.3 translates to an absolute magnitude limit of -14.6 and -16.8 at the redshifts of 0.13 and 0.32, respectively.

\begin{figure*}
    \centering
    \includegraphics[height=0.5\textwidth]{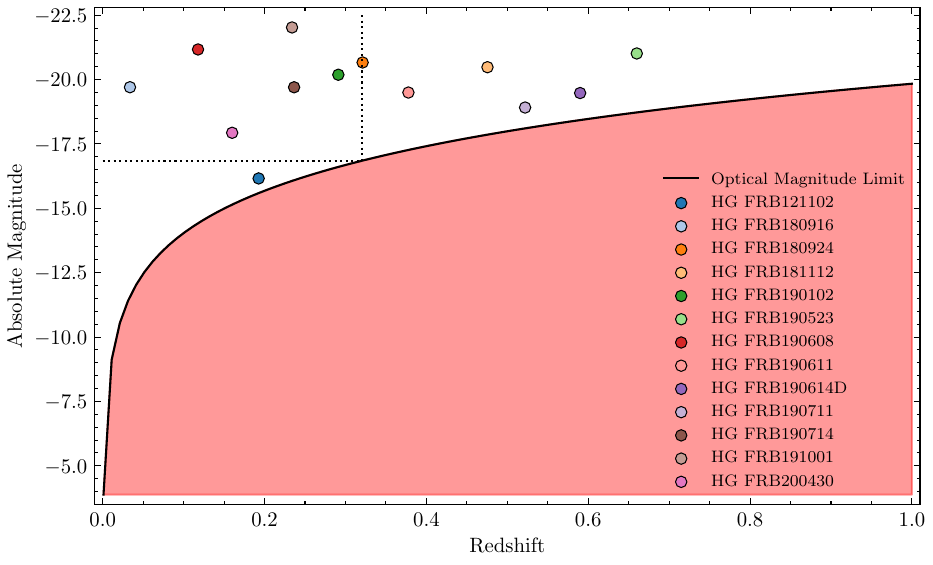}
    \caption{The black solid line is the optical absolute magnitude limit (corresponding to our Gemini $r$-band apparent magnitude limit of 24.3) with respect to redshift. Absolute magnitudes of host galaxies of localized FRBs are shown with colored circles. The black dotted line represents the absolute magnitude limit of our observations at the upper limit of redshift (z = 0.32) for \src. Sources within the red shaded region would not be detected with our observations.}
    \label{fig:opt_plot}
\end{figure*}

Figure~\ref{fig:opt_plot} shows the optical limit of our observations as a function of redshift. It also shows absolute magnitudes of host galaxies of localized FRBs with respect to their redshifts \citep{Macquart2020, bannister2019, Ravi+2019, bhandari2020, law2020, Prochaska2019Sci, chatterjee17, Heintz2020}. The magnitude limits at these redshifts are higher than those of host galaxies of localized FRBs. Therefore, if the host galaxy of \src\ is similar to that of other localized FRBs, then we would have detected it: the host galaxy would be one of the galaxies in Table~\ref{tab:src_loc}.   

%Our optical observations were sensitive to detect all host galaxies, except that of FRB121102. We would also have detected these host galaxies, except that of FRB121102 and FRB200430, if they were at the upper limit of the redshift estimate of \src\ (i.e z = 0.51, see Fig~\ref{fig:opt_plot}). 
Our observations were also complete to all Milky Way type galaxies ($M\sim-21$) and all bright elliptical galaxies within the redshift upper limit for \src. 

Thus, it seems likely that we have narrowed down the host to one of those listed in Table~\ref{tab:src_loc}. However, a caveat to this analysis is that our observations were not complete to low-surface brightness galaxies \citep{lsb_review}, but thus far, no FRB has been localized or associated to these galaxies.

%Lack of a coincident association thus disfavors the plausible host of \src\ to be a \fixme{blah} galaxy. 

%We have plotted the luminosity limits of our observations with respect to redshift in Fig~\ref{fig:lum_plot}). The horizontal lines represent the upper limits on the PRS luminosity for all the localized FRBs, and the PRS luminosity of FRB\,121102. 

%\fixme{Make some comment about how this compares with the luminosity of FRB\,121102's PRS. If this FRB had a PRS like FRB\,121102's and was at those redshifts, would we have seen the PRS?}

%Considering the non-detection of any associated source with the FRB location, we can place a $3\sigma$ upper limit of 17.1 $\mu Jy$ on the flux of any persistent radio source. Using the redshift upper limit of 0.6 (see section~\ref{sec:redshift}), this corresponds to a luminosity of $< 3 \times 10^{38}$ erg/s.

%The luminosity of the brightest radio source within the localization region, with a peak flux of \fixme{22.1 $\mu$Jy}, at its maximum photometric redshift estimate is \textbf{blah}.  

\subsection{Luminosity of the FRB}
\citet[][]{oslowski2019} estimated the fluence of \src\ to be $>83.5$\,Jy\,ms 
%from its S/N in the data from the pulsar instrument 
(see their section 4.1 for details). 
%Using their reported width of 0.475\,ms, 
This translates to a luminosity of $>8.7\times 10^{32}$\,erg\,Hz$^{-1}$ %$>6.0 \times 10^{28}$\,W\,Hz$^{-1}$ %$>1.8 \times 10^{29}$\,W\,Hz$^{-1}$ 
at the redshift of 0.32 (the redshift upper limit for \src). We can compare this with the luminosities of all localized FRBs \citep[][]{Macquart2020, marcote20, law2020, bhandari2020, bannister2019, Ravi+2019, Prochaska2019Sci, chatterjee17}.
%with a flux and DM reported in the FRB Catalogue\footnote{\url{http://www.frbcat.org/}} \citep{Petroff2016}, assuming that all of the FRBs are at the maximum distance inferred by their DMs. 
Within this sample, we find that \src\ is the second most luminous FRB (with one-sixth the luminosity of the most luminous localized FRB, FRB\,190523). However, it may be much brighter given we have only a lower limit on its fluence and an upper limit on the redshift. Moreover, it is likely that FRBs exhibit beamed emission which introduces an inherent uncertainty in comparing observed fluences.
%and is the brightest FRB detected by Parkes Telescope.   

%\fixme{I feel like one big oversight here is that we try to estimate how far the FRB is and have some range in redshifts and potential hosts but we don't have any mention of the implications of the energetics involved in this very high-S/N FRB.}

\section{Conclusion}
We report the results of multi-wavelength follow-up observations of \src\ to search for a host galaxy and non-bursting persistent radio emission. \src\ was detected by Parkes Telescope during PPTA observations and is one of the brightest FRB detected with the Parkes Telescope to date. We used the multi-beam detection of this FRB to improve the localization precision to a $\sim2'\times 2'$ region, a factor of $\sim$37 improvement when compared with a typical localization with the Parkes Telescope. We estimated the redshift of this FRB to lie in the range 0.13--0.32. %using two independent approaches. 
In our Gemini observations, we identified 20 galaxies within the localization error region. We found that the error region is still sufficiently large---and the galaxies' magnitudes sufficiently faint---that we could not confidently associate 
any of these sources with \src. 
Two of these galaxies also had a potential co-located VLA radio detection, however, these serve as only weak candidates for any persistent radio emission. No other coincident radio sources were found above the $3\sigma$ limit of 17.1$\mu$Jy\,beam$^{-1}$.  If a putative PRS of \src\ were similar to that of FRB\,121102, we would have detected it out to the farthest likely redshift of \src. We did not detect any time-varying or appropriately redshifted \hi\ feature in our ATCA observations. 

Far from our localization region, but within our radio imaging field, we detected a complex X-shaped source (\xsrc). The morphology of this source is similar to the S-shaped sources reported in the literature. 

Ultimately, a comparison of the photometric redshifts of our 20 candidate host galaxies with the DM-based redshift range of \src\ resulted in 14 galaxies remaining as the most likely hosts of this FRB. While it is beyond the scope of this work, follow-up spectroscopy and in-depth analysis of the ionization content (i.\,e. potential for host DM contribution) for each of these galaxies may help reveal which galaxy is the host of this luminous FRB. Moreover, it might not be possible to unambiguously identify its host galaxy unless this FRB is detected to repeat and localized by an interferometer.
%\fixme{If appropriate here, note limitations, which is that there may well be a host that we didn't identify because our magnitude limits weren't deep enough.}

%\fixme{Add paragraph here about what's needed for the next steps... e.g. spectroscopy and SFR analysis of the galaxies we reported here. Note limitations, which is that there may well be a host that we didn't identify because our magnitude limits weren't deep enough.}

%We discuss this source within the framework of proposed models for X-shaped sources. \verify{We believe that this source could be a result of reactivation of jets in an AGN due to external interaction, or could also be explained by dual-AGNs with misaligned jets.} We recommend deeper optical and radio observations of this source to further understand its morphology. 

\section*{Acknowledgements}
KA and SBS acknowledge support from NSF grant AAG-1714897.  
N.T. acknowledges support by FONDECYT grant 11191217. G.P acknowledges support by the Millennium Science Initiative ICN12\_009.
Based on observations obtained at the international Gemini Observatory, a program of NSF’s OIR Lab, which is managed by the Association of Universities for Research in Astronomy (AURA) under a cooperative agreement with the National Science Foundation, on behalf of the Gemini Observatory partnership: the National Science Foundation (United States), National Research Council (Canada), Agencia Nacional de Investigaci\'{o}n y Desarrollo (Chile), Ministerio de Ciencia, Tecnolog\'{i}a e Innovaci\'{o}n (Argentina), Minist\'{e}rio da Ci\^{e}ncia, Tecnologia, Inova\c{c}\~{o}es e Comunica\c{c}\~{o}es (Brazil), and Korea Astronomy and Space Science Institute (Republic of Korea). The Gemini data was obtained from program GS-2018A-Q-205 and processed using the Gemini {\sc Pyraf}
package\footnote{\url{https://www.gemini.edu/sciops/data-and-results/processing-software/}} \citep[][]{pyraf}. The National Radio Astronomy Observatory is a facility of the National Science Foundation operated under cooperative agreement by Associated Universities, Inc. The Common Astronomy Software Applications (CASA) package is software produced and maintained by NRAO. The Australia Telescope Compact Array (ATCA) is part of the Australia Telescope National Facility which is funded by the Australian Government for operation as a National Facility managed by CSIRO.  We acknowledge the Gomeroi people as the traditional owners of the ATCA site. The Parkes radio telescope is part of the Australia Telescope National Facility which is funded by the Australian Government for operation as a National Facility managed by CSIRO. We acknowledge the Wiradjuri people as the traditional owners of the Observatory site.

\facilities{EVLA, Gemini, ATCA}

\software{astropy \citep{astropy:2013, astropy:2018}, karma \citep{karma}, CASA \citep{casa}, numpy \citep{numpy}, matplotlib \citep{Hunter:2007}, pandas \citep{pandas}}

\appendix
\section{Association Probability Calculation}
\label{appendix:assoc}
The chance association probability of an FRB to a host galaxy depends significantly on the surface density of the galaxies on the sky, offset of the source from the galaxy and localisation uncertainties. Following \citet[hereafter B02]{bloom2002} and \citet[hereafter EB17]{eftekhari2017} we calculated the localization region (R) using $R = \text{max}[2R_{FRB}, \sqrt{R_0^2 + 4R_h^2}]$, where $R_{FRB}$ is the 1$\sigma$ localization radius of the FRB, $R_0$ is the radial angular separation between the FRB position and a presumed host, and $R_h$ is the galaxy half-light radius. We use typical values of $R_0$ and $R_h$ for LGRB and SLSN host galaxies, as given in \citetalias{eftekhari2017}. 

For the chance probability calculation, we followed Section 2 of \citetalias{eftekhari2017}. We did a 3$^{\text{rd}}$ order spline fit to the r-band galaxy number counts given in Table 3 of \citet{driver2016}. In cases where multiple number counts were present corresponding to the same magnitude bin, we chose the one with the least cosmic variance. We did not weight our magnitude bins using cosmic variance for the spline fit. The probability of chance coincidence was then defined as in equation 1 of \citetalias{eftekhari2017}. 

Further, we also followed Section 3 of \citetalias{eftekhari2017}, to use the redshift constraint on the FRB to estimate the chance association probability. Here, relationships between DM and $z$ are used to estimate the likely range in redshift of the FRB from its DM \citep{Macquart2020, pol2019, ioka2003, inoue2004}. We calculate the number density of galaxies by integrating optical luminosity functions presented in \citet[][Table 6, blue galaxies]{Beare2015} for 0.2 $<$ z $<$ 1.2 and \citet[][]{blanton2003} for z $< 0.1$. To estimate the luminosity function for 0.1 $<$ z $<$ 0.2, we averaged the luminosity function parameters for z $<$ 0.1 and 0.2 $<$ z $<$ 0.4 (T. Eftekhari private communication). %The definition of probability given in equation 2 in \citetalias{eftekhari2017} was used here. 
The chance association probability is then given by: 
\begin{equation}
    P_{cc} = 1 - e^{-f_A N(\leq M,\leq z)}
\end{equation}
Here, $f_A$ = $\pi R^2/5.346\times 10^{11}$ is the fractional area of the localization region on the sky, where $R$ is in arcseconds. $N(\leq M,\leq z)$ is the total number of galaxies above a limiting absolute magnitude M, within a comoving volume out to a redshift z. As luminosity functions are different for different redshift bins, we calculated the number of galaxies in each redshift bin individually. The number of galaxies in each bin is calculated by integrating the luminosity functions from absolute magnitude of $-24$ to absolute magnitude corresponding to 0.01L* galaxy, multiplied by the comoving volume of that redshift annuli. The total number of galaxies was obtained by adding the number of galaxies from the lowest redshift bin (i.e. 0 $<$ z $<$ 0.1) to that including max redshift (z$_{max}$) bin. 

We also provide \texttt{casp}\footnote{\href{https://github.com/KshitijAggarwal/casp}{https://github.com/KshitijAggarwal/casp}}: Calculating ASsociation Probability of FRBs, which is a python package \citep[][]{casp} and a webpage\footnote{\href{https://kshitijaggarwal.github.io/casp/}{https://kshitijaggarwal.github.io/casp/}} to calculate association probability of FRBs using \citetalias{bloom2002, eftekhari2017}.   

\bibliography{frbfollowup}

\end{document}